# *Flooding through the lens of mobile phone activity*

David Pastor-Escuredo[1], Alfredo Morales-Guzmán[1], Yolanda Torres-Fernández[1], Jean-Martin Bauer[2], Amit Wadhwa[2], Carlos Castro-Correa[3], Liudmyla Romanoff[4], Jong Gun Lee[4], Alex Rutherford[4], Vanessa Frias-Martinez[5], Nuria Oliver[6], Enrique Frias-Martinez[6], Miguel Luengo-Oroz[4]

[1]Universidad Politécnica de Madrid [2]Vulnerability Analysis and Mapping, World Food Programme [3]Coordinación de Estrategia Digital Nacional, Presidencia de la República de México [4]United Nations Global Pulse [5]University Maryland [6]Telefónica Research

*Abstract*—Natural disasters affect hundreds of millions of people worldwide every year. Emergency response efforts depend upon the availability of timely information, such as information concerning the movements of affected populations. The analysis of aggregated and anonymized Call Detail Records (CDR) captured from the mobile phone infrastructure provides new possibilities to characterize human behavior during critical events. In this work, we investigate the viability of using CDR data combined with other sources of information to characterize the floods that occurred in Tabasco, Mexico in 2009. An impact map has been reconstructed using Landsat-7 images to identify the floods. Within this frame, the underlying communication activity signals in the CDR data have been analyzed and compared against rainfall levels extracted from data of the NASA-TRMM project. The variations in the number of active phones connected to each cell tower reveal abnormal activity patterns in the most affected locations during and after the floods that could be used as signatures of the floods - both in terms of infrastructure impact assessment and population information awareness. The representativeness of the analysis has been assessed using census data and civil protection records. While a more extensive validation is required, these early results suggest high potential in using cell tower activity information to improve early warning and emergency management mechanisms.

## I. Introduction

Natural disasters such as floods or earthquakes affect hundreds of millions of people worldwide every year[1]. The effectiveness of humanitarian response is limited, in part, due to the lack of timely and accurate information about the patterns of movement and communication of affected populations. Specifically, there is a need for dynamic in-situ information across the event timeline: a baseline understanding of *regular behavior* before the onset of an emergency, real-time information about the behavior of a disaster-affected population, and the capacity to track return to normal behavioral patterns during the recovery phase. Governments, international organizations and humanitarian actors could potentially enhance the effectiveness of their response by gaining access to accurate geospatial and temporal information of population displacements and communication patterns before, during and after a disaster occurs.

Over the last few years, due to the exponential increase in the penetration of mobile phones, new opportunities for obtaining such indicators have emerged. In particular, the use of mobile phones as sensors of human behavior has yielded important research findings in large-scale social dynamics analysis in areas such as human mobility, information diffusion, social development, epidemiology and disaster response. A commonly used source of mobile phone data for these studies are aggregated and anonymized Call Detail Records (CDRs), which provide data about phone activity within a mobile network and are described in Section III.B. In the area of human mobility, various approaches have shown the viability of using CDR data to model mobility patterns in both developed and developing economies [1][2][3] and also the impact of population mobility during disease outbreaks [4]. Various studies have redefined our understanding of information propagation [5][6] to characterize cooperative human actions under external perturbations and have offered new perspectives on panic [7][8][9][10].

In the area of social development, CDR analysis has also shown promise to understand migration patterns in urban settlements (slums) in Kenya, enabling researchers to infer informal employment [11]; to infer demographic [12] and socioeconomic information in developing countries in Latin America [13]; or to characterize population movements [14]. Finally, CDR data has also been successfully applied to model and evaluate natural disasters. A study after an earthquake in Haiti found that a CDR-based estimation of population movements during disasters or disease outbreaks can be delivered rapidly and with high accuracy [15]. Similar studies using CDR data also showed the ability to measure the impact of earthquakes on communication patterns [16] and to build predictive models of areas of disruption following an earthquake [17]. In general, CDRs are expected to contain different signatures –spatial and temporal patterns– of social behavior during different type of events and emergencies [18] that could be used for early response. Moreover, mobile phones can also be used as sensors to obtain other data besides social variables, such as precipitation measurements [19].

---

[1] EM-DAT database: http://emdat.be/disaster-trends

In this work, we are interested in exploring potential signatures implicit in CDR data as a means for characterizing real phenomena taking place during floods. These studies could one day be applied in ways that reduce mortality and improve outcomes for disaster-affected populations.

## II. Objectives

The objective of this research is to develop and apply methods to assess the suitability of using aggregated and anonymized CDR data to characterize the impact of floods on populations, using the Tabasco, Mexico floods in 2009 as a case study. Our ultimate goal is to contribute to the development of real-time CDR based decision-support tools for public sector response to floods and other natural disasters.

The technical contributions of this work are (1) a multimodal data integration framework that facilitates the integration of CDR data with other data sources- remote sensing, rainfall activity, census and civil protection information and (2) the quantitative characterization of changes in communication patterns during the floods and their relation to external ground truth information.

## III. Problem description

*A. Context: Tabasco floods in 2009*

The state of Tabasco is located to the south of the Gulf of Mexico, covering 24,738 sqkm (1,3% of national total area). Due to its location and topographical features, Tabasco is subject to frequent flooding events, such as those that occurred in 2007, 2008 and 2009. On 28th October 2009, a cold front Nr. 9 entered northwest Mexico and reached Tabasco on the 31st, where it remained for four days. It rained intensely until November the 3rd over the west of Tabasco, within the Tonala basin. The National Meteorological Service (SMN) recorded 800mm of cumulated rain in three days, 4-fold the regular cumulated rain level for November. Due to these figures, the precipitation was classified as extraordinary.

As the Tonala basin lacks hydraulic infrastructure for controlling river floods, the rain water flowed freely to the coastal plains, causing flooding. The greatest damage occurred in the Huimanguillo and Cardenas municipalities. On November the 3rd, after the heavy rain, the state of emergency was declared in Huimanguillo and Cardenas. Response activities coordinated by Civil Protection and the system for Integral Development of Families (DIF), with contributions from other state and federal entities, such as the Federal Preventive Police and the National Water Commission (CONAGUA). On November the 11th, a state of emergency was declared in Comalcalco, Cunduacán, and Paraíso municipalities.

In January 2010, the National Center for Disaster Prevention (CENAPRED) carried out a mission to assess the damage caused by the floods, together with the Planificación State Secretariat and Civil Protection. They interviewed over 16 state and federal agents in charge of coordinating recovery actions. CENAPRED collected all the information and compiled a report on the impact of the floods. According to the report, in economic terms, the total losses in the state of Tabasco reached 190 million USD, 50% of which were due to damage to road infrastructure (see Fig.1); 16% were related to productive activities (agriculture and ranching); and 7% of losses corresponded to social damage (dwelling, health, education). The floods also had a significant emotional and psychological impact on people's lives.

The CENAPRED report states that the total human, social and economic losses caused by the 2007, 2008 and 2009 stationary floods highlight the vulnerability of Tabasco to such natural events. Furthermore, this recurring situation hinders the state from achieving total recovery after each disaster. Hence it is recommended that resources be invested in designing and implementing mitigation plans and prevention actions rather than in covering post-event costs.

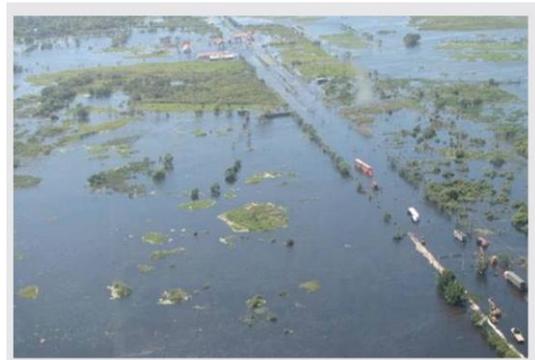

*Fig.1: Federal road 180D totally submerged. Transit problems complicated evacuation and emergency aid activities. Source: CENAPRED*

*B. Call Detail Records (CDRs)*

Cell phone networks are built using a set of base transceiver stations (BTS) that facilitate communication between cell phone devices within the network. Each BTS has a geographical location represented by its latitude and longitude. The area covered by a BTS is called a cell, and can be approximated using Voronoi tessellation. At any given moment, one or more BTSs can provide coverage to a cell phone. The final BTS is assigned depending on the network traffic and on the geographic position of the phone.

CDR (Call Detail Record) databases are generated when a mobile phone connected to the network makes or receives a phone call or uses a service (e.g., SMS, MMS, etc.). In the process, and for billing purposes, the information regarding the time of the event and the BTS tower that the phone was connected to the even occurred is logged, which gives an indication of the coarse geographical position of the phone at a given moment in time; no precise position of the phone is recorded or calculated.

Among all the data contained in a CDR, our study uses the anonymized (encrypted) originating number, the anonymized (encrypted) destination number, the time and date of the call, the duration of the call, and the latitude and longitude of the BTS used by the originating cell phone number and the destination phone number when the interaction happened. The dataset available for this study contained only CDRs generated by the BTSs contained in the geographical area affected by the floods (roughly the state of Tabasco and parts of Veracruz). All the data was not only anonymized but also aggregated. No contract or personal data was collected, accessed or utilized for this study. No authors of this study participated in the extraction of the dataset.

*C. Additional data sources*

Additional data sources analyzed in this research include:

*a) Satellite imagery data*

Multispectral, medium resolution (15 to 60 meters) ETM+ Landsat7[2] satellite images have been used for detecting and delimitating the submerged land. The temporal resolution of this data source is 16 days, so it helps to approximate the flooded area with reasonable accuracy, at least before and after the flooding happened. The spatial resolution is high enough to segment broad floods, river overflows or lake leakages. The satellite imagery data allows us to spatially limit the affected regions with better accuracy than the vague approximations that could be inferred retroactivitly from news or historical documents.

*b) Precipitation data*

The Tropical Rainfall Measuring Mission project[3] provides high resolution (3 hours of temporal resolution and 0.25 squared degrees of spatial resolution) of precipitation levels worldwide. The spatial resolution of this data is lower than the satellite images used to segment the floods, but high enough to obtain a realistic precipitation level in the affected area. On the other hand, the temporal resolution is adequate to generate a time series comparable to the CDR data.

*c) Civil protection data*

Once the state of emergency was declared, Tabasco Civil Protection assisted affected people in Huimanguillo and Cardenas first, and in Comalcalco, Cunduacán, and Paraíso days later. People who directly suffered the effects of the floods were moved to emergency camps and received first aid and staple goods (like water, food, blankets). Civil Protection recorded the data from Table 1, which we have used to validate the results obtained from the other data sources.

*d) Census data*

The most recent official Census[4] of Mexico (2010) has been used to assess the representativeness and validate the population distribution inferred with the CDR data.

*Table 1. Affected population and emergency camps in several municipalities of Tabasco. Source: SEGOB.*

| Municipality | Affected population | Nr. of temporal camps |
|---|---|---|
| Cárdenas | 105,272 | 69 |
| Comalcalco | 18,215 | 15 |
| Cunduacán | 10,280 | 5 |
| Huimanguillo | 53,688 | 60 |
| Jalpa de Méndez | 147 | 0 |
| Paraíso | 27,134 | 14 |
| **Total** | **214,736** | **163** |

*e) The Global Administrative Areas Database (GADM)*

The GADM[5] provides GIS-compatible maps of administrative areas worldwide. GADM was used to classify the antennas locations in the map and associate them to the administrative boundaries of the state of Tabasco.

*f) Other contextual information*

Diverse data sources were consulted in order to get a wider understanding of the situation:

- The *Tropical Cyclones Early Warning System* (SIAT CT) from the Mexican Civil Protection website[6]. In this document, the different phases of a tropical storm are clearly explained, as well as the actions designed by Civil Protection to respond to each phase. The actions are detailed chronologically in the emergency plan. We used this information to define the time scale for the temporal analysis of CDRs and precipitation data, to be later correlated with the population's behavior patterns.

- *Flood hazard, vulnerability and risk maps* from the National Center for Disaster Prevention[7], were used to become acquainted with the prevention and mitigation flood risk studies carried out in the country.

- *News and photos* about the consequences of the floods from local digital newspapers and blogs, such as *El Economista*[8], *La Jornada*[9], *Informador*[10], among others. We geo-located relevant events like injured people, damaged infrastructure, river overflows and isolated towns, in order to gain a preliminary sense of the affected areas and the spatial distribution of damages.

---

[2] http://earthexplorer.usgs.gov/
[3] http:// http://trmm.gsfc.nasa.gov/
[4] http://www.censo2010.org.mx/
[5] http://www.gadm.org/
[6] http://www.proteccioncivil.gob.mx/work/models/ProteccionCivil/Resource/62/1/images/siatct.pdf
[7] http://www.atlasnacionalderiesgos.gob.mx/index.php/biblioteca/category/17-hidrometeorologicos
[8] http://eleconomista.com.mx/politica/2009/11/08/inundaciones-tabasco-suman-200000-damnificados
[9] http://www.jornada.unam.mx/2009/11/11/estados/034n2est
[10] http://www.informador.com.mx/mexico/2009/151290/6/inundacion-deja-siete-mil-115-damnificados-en-tabasco.htm

## IV. METHODOLOGY

The methodological framework proposed in this study comprises three main steps (see Fig.2): (1) *Evaluation of the Representativeness of the Data:* In order to study social behavioral patterns within CDR data it is necessary to evaluate how representative the mobile subscribers within it are of the target populations. We used the 2010 census of Tabasco as the ground truth to measure the representativeness of the signals extracted from the CDR data depicted in III.B; (2) *Data Integration:* additional data sources described in III.C have been gathered, homogenized and integrated into a Geographic Information System (GIS) creating a geo-spatial frame enabling interpretations of the CDR data analysis. The CDR data serves as the higher resolution substrate in which we integrate other independent data with different spatial and temporal resolutions;

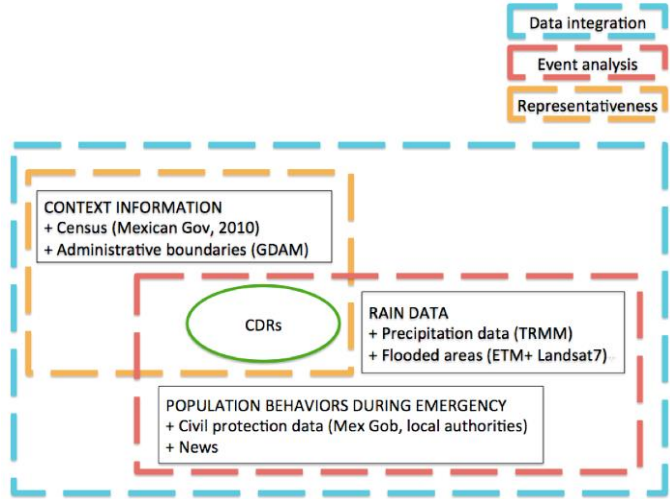

*Fig.2: Overview of the methodological framework and data sources*

(3) *Data-driven Event Analysis:* In order to perform the analysis of CDR and remote sensing data, we have developed custom processing methods. For the CDR data analysis, a library of tools in Python has been implemented to parse CDR database files, filter them according to their associated GPS coordinates, reconstruct displacement trajectories, measure statistical descriptors, and visualize them together with geo-referenced data. In order to analyze remote sensing images and identify the flooded area we have implemented an image processing pipeline that uses mathematical morphology (see Fig.3) and a maximum likelihood per-pixel classification method --available in ArcGIS software—to detect small water concentrations and to refine the boundaries of the wider previously segmented regions. The raw precipitation data described in III.C.b has been analyzed with MATLAB and Python scripts. A GPS conversion transformation has been applied to retrieve precipitation data at the antenna position (see Fig.4). As expected, the accumulated rainfall information matches with the segmented floods with the methodology described above.

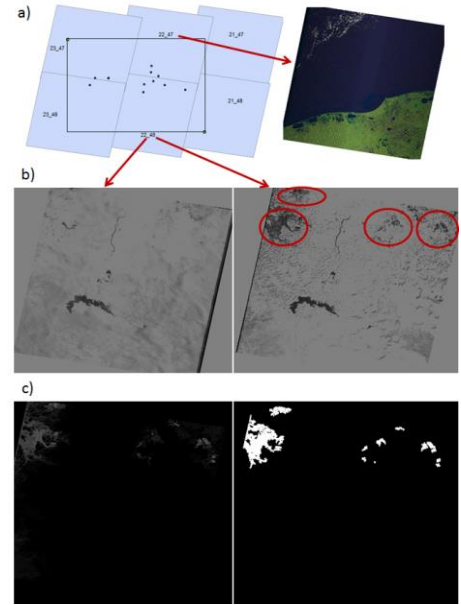

*Fig.3: a) Six panels of the data described in III.C.a were needed to cover Tabasco. b) A set of images pre-floods was used as a reference for comparison to another set of images obtained right after the floods in order to identify floods. c) Gaussian filtering and morphological geodesic reconstruction from seeds were used to semi-automatically segment flooded areas.*

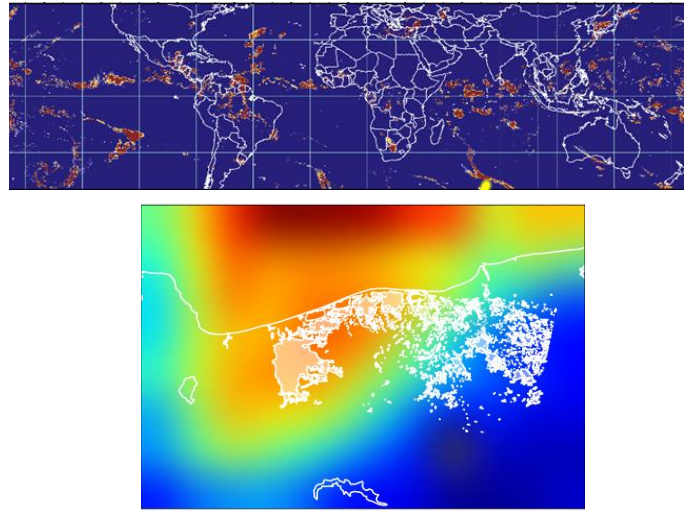

*Fig.4: Top: TRMM project raw data [-180,-50] to [180,50] degrees at one-day accumulation resolution (Nov 1$^{st}$). Bottom: Accumulated precipitation during the first two weeks of November in Tabasco overlaid with the segmentation using satellite imaging.*

## V. RESULTS

### A. Assessing the representativeness of CDR data

In this study, we considered a subset of the CDRs comprising only the phone calls (social baseline) made from Tabasco during the month prior to the onset of the reported floods on November 1st, 2009 (baseline period). Figure 6 shows the spatial distribution of the antennas covering the study area. In order to evaluate how representative this data is of the real population of Tabasco, we have compared the population distribution derived from the antenna activity with the 2010 census of Tabasco, used as the ground truth. The underlying hypothesis here is that CDR- based analysis may be extrapolated to measurements over the full population if the subscribers are homogeneously distributed compared to the real census, provided that the sampling of CDR data in the region is sufficient.

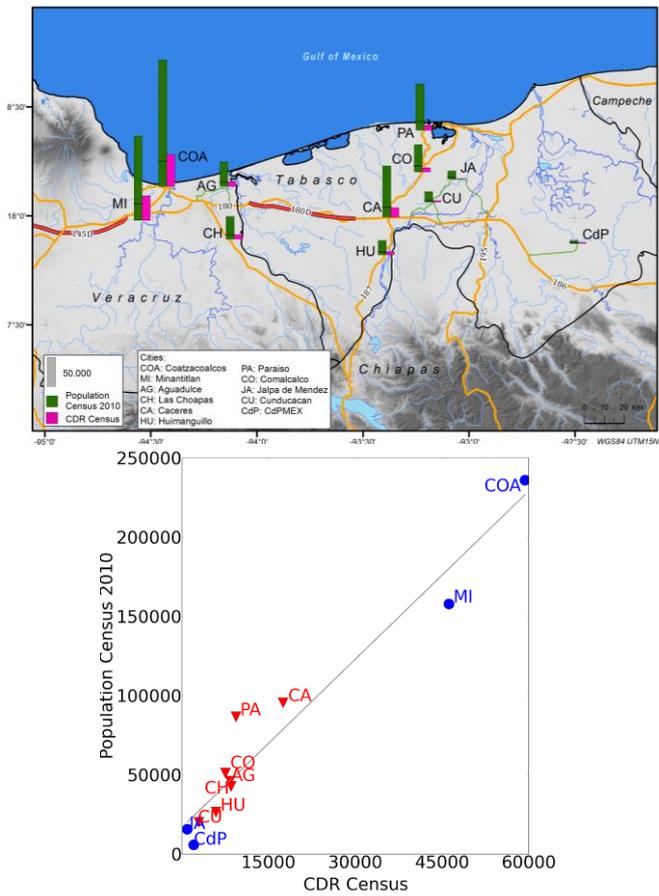

*Fig.5: Top: map of 2010 census (green bars) vs CDRs based population estimation (purple bars) in several cities of Tabasco (red=affected cities, blue=other cities) and surroundings. Bottom: The plot shows linear correlation between the CDR census and the real census (r-square 0.97).*

The social baseline has been characterized by assigning the *home antenna tower (HAT)* for each phone, meaning the antenna tower most used at night during the baseline (BL) period [21]. The number of users per city (or administrative boundary) was inferred by cross-referencing the users' HAT with the GADM database. We then compared the 2010 census information with the CDR population estimation for the main cities of the regions affected by the 2009 floods: Cárdenas, Huimanguillo, Paraíso, Comalco, Cunduacán and other nearby cities (see Fig.5). Results showed a linear relation between both variables with a relative homogeneity of around the 20%. Hence, this analysis provides preliminary results that support the assumption of a homogeneous representativeness of the CDR-based data in the affected cities, enabling us to use the proposed hypothesis in this study.

### B. Population response to the floods based on variations in cell tower activity

For the analysis, the CDR data of the baseline has been aggregated by day and by antenna to understand how the floods modulated the normal communication patterns observed at the antenna level [18]. In particular, we measured the number of unique phones placing or receiving calls in each antenna and for each day. We refer to this raw measurement as the *BTS communication activity x(t)* (see Fig.6 Top)

To detect abnormalities in this activity, such as those produced by the floods, we propose the *BTS variation* metric that relies on the comparison $x(t)$ against their characteristic variation obtained during the baseline period. Mathematically, the BTS variation metric -$x_{norm}(t)$- is defined as the z-score from $x(t)$ referred to the normal distribution characterizing the baseline pattern as follows:

$$x_{norm}(t) = \frac{x(t) - \mu_{BL}}{\sigma_{BL}}$$

where the pair $(\mu_{BL}, \sigma_{BL})$ statistically characterizes the activity during the BL period (the month before the flooding onset). A static z-score has been previously used to characterize calling behaviors in large scale time sensitive emergency events like bombings, earthquakes or brief storms [18]. Here, we have computed $x_{norm}(t)$ from the beginning of the BL period until the end of January (~2 months later the rainfall finished), generating temporal series of this z-score for the BTSs in the areas affected as shown in the Appendix-Figure. The spatial distribution of the maximum value of the BTS variation metric $x_{norm}(t)$ -derived from the CDRs- is shown in an impact map (see Fig.6 and Appendix-Video) that combines the metric with other contextual indicators: the municipalities have been colored according to the official number of affected population and the segmentation of the flooded area generated from the Landsat-7 images. The impact map is consistent with our ground truth evidence (flood segmentation and civil protection records), since the BTS activity spikes in the most affected municipalities: Cárdenas and Huimanguillo (Fig.6 Bottom).

The BTSs featuring high variations of the metric outside of the affected regions are mainly those BTSs located near the ground transportation system. This might be a useful indicator for resource allocation in future emergencies. For example, very high variations are observed along Federal Road 180D, which was eventually completely covered by water (see Fig.1). Note that the temporal series of the Appendix-Figure also shows strong variations during a segment representing Christmas, where most of the sampled BTSs in the region spike, whereas the floods only trigger changes in the nearby BTSs.

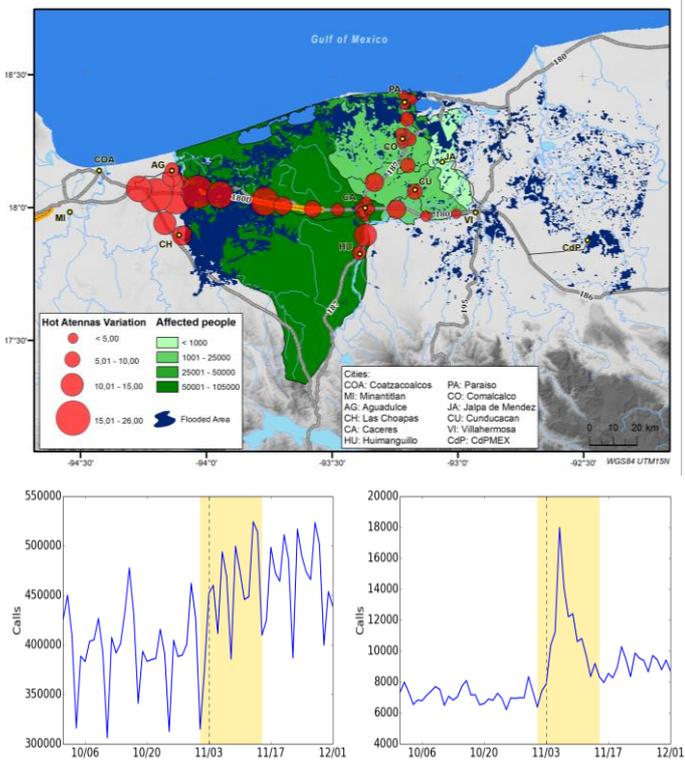

Fig.6: Top: "Impact Map" of Tabasco for the 2009 floods. The map shows the most critical day featuring the highest values of the BTS variation metric. See Appendix-Figure for the temporal series of these BTSs. Bottom: Signal $x(t)$ aggregated for all antennas in Tabasco (left) and for the antennas close to the floods segmentation (right).

During the floods, the distribution of the maximum in the BTS variation metric is wider than the BL period distribution, featuring more BTS with higher variation metric (see Fig.7).

The real-time nature of mobile phone signals allows us to compare social patterns against their modulating factors. Here, we compare the proposed metric with rainfall levels. These precipitation levels are obtained from the NASA TRMM project's day-resolution estimations of the rainfalls. The six hottest BTS that also feature different metric profile have been taken to observe the rainfall levels at the BTS level (see Fig.8 Top). As shown, the typical delay between the maximum level of precipitations and the peak in the variations of the hot BTS indicator is 4 days. One possible explanation is that a population might not react in a way that alters the communication activity globally even under extreme climatological conditions. Instead, the response captured in the communication activity could have occurred due to the initial flooding effects, after the rivers and water reserves overflowed around November 5th and 6th as was reported in different news (see section III.C.f).

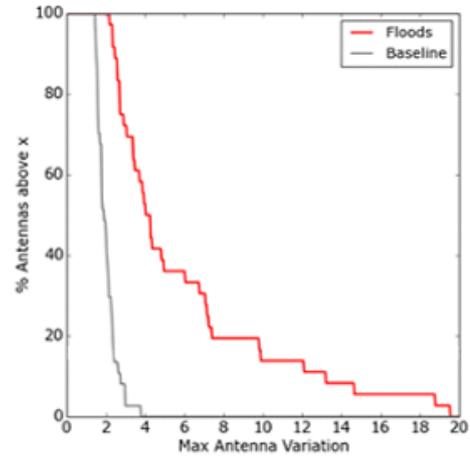

Fig.7: Distribution of the maximum of the BTS variation metric for the BL period (gray) and floods (red). The curves show the percentage of antennas (y-axis) whose maximum metric value is higher than a given value (x-axis).

The civil protection warning was issued on the day of maximum precipitations (November 3rd). It would be expected that this warning would result in a spike in communications activity, but this reaction can only be observed in two BTSs located along Federal Road 180D that eventually suffered an outage (see Fig.8 Bottom). These sudden variations and the following outage may indicate the point of the highest rain impact, likely causing a severe traffic jam on 180D. The increase of the BTS occupancy time due to the jam would eventually generate the shown communication activity peaks (although further analysis would be required).

On the other hand, the maximum of the BTS variation in the antennas with higher population happens on November 6th when the rain was already vanishing. Several sources also raised the estimates of the affected population from 50,000 to 100,000 people that day. Thus, the hypothesis would be that for gradual-onset disasters (due to a cumulative effect of some potential factor), the proposed metric might provide an estimation of the population's awareness and subsequent reaction rather than a means to detect the onset of the event. The delayed spike in BTS variation in this case may indicate that while the civil protection warning did not produce the sufficient level of awareness in the population, the initial consequences of the flooding did.

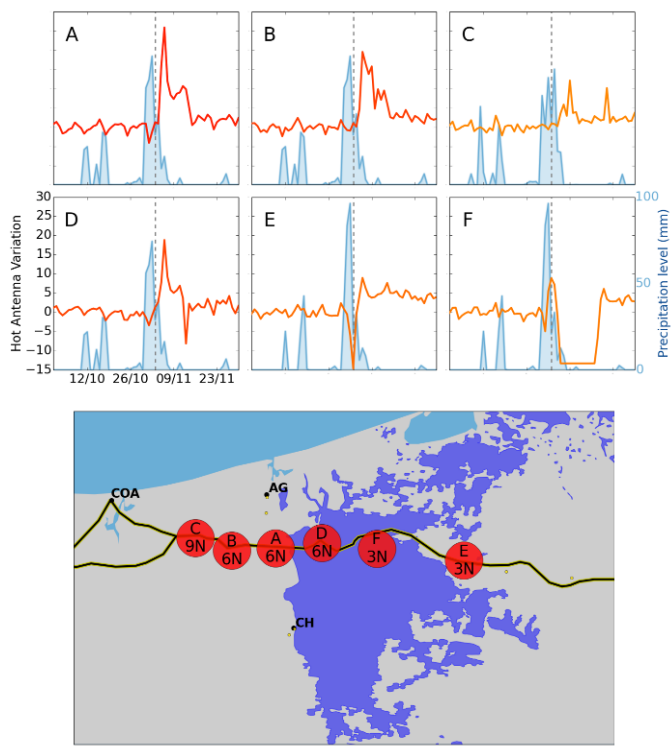

*Fig.8: Top: BTS variation metric (red) vs the precipitation level (blue) for the six hottest BTS. The slashed line shows the emergency warning date as notified in the news. Bottom: Map featuring the position and date (e.g. 6N is 6$^{th}$ November) where the maximum of the BTS variation metric was observed.*

## VI. DISCUSSION

This work is a retrospective study that leverages the footprints of mobile phone activity during floods to propose data-driven indicators with potential to support decision making during emergencies. In particular, we have proposed a methodology based on integrated analysis of CDRs with several data sources, including remote sensing imagery and rainfall information. We have first tested the representativeness of the CDR data, observing a homogeneous penetration of mobile phones in the affected cities enabling us to use the hypothesis that CDR-based analysis may be extrapolated to estimate measurements over the full population. Therefore, it would be possible to estimate population changes in regions with sufficient density of BTSs (as remote sensors of communication activity), provided that changes in the size of the population outpace changes in mobile carrier penetration.

We also tested a CDR-based measurement to discover abnormal communication patterns at the cell tower level. The information on abnormal cell tower activity as a result of floods could be used to trigger further investigations and to potentially locate damaged areas, assess needs and allocate resources in the short term (for example sending additional supplies to nearby centers). In particular, this would allow improved resource allocation in the first 24-48 hours when resources are scarce. The identification of relevant affected cell towers might also serve to better target public communications. Our findings show that results are relevant at the BTS-level. While the data is fully anonymized and aggregated, changes in the activity in the cell towers proves of utility for emergency operations.

Abnormal communication activity might also be used to measure the awareness of at-risk populations indicating those insufficiently responsive to early warning announcements. Note that the population's reaction --in terms of increased communication and hence increased activity in the cell towers—took place when the emergency was declared, rather than during the previous alert stage, as expected. This could be an indicator of the skepticism or lack of awareness of the population regarding the heightened risk of floods. If this is the case, a systematic study of the reasons for such behavior is recommended, since a lack of awareness of the existence of a hazard implies an increase in vulnerability to its effects. This study suggests that, in the future, the citizens' reactions to a catastrophe during the emergency phase could be assessed and incorporated into an evolving emergency management strategy.

Note that the proposed indicators are candidates for further exploration; these methods ideally need fine-grained validation with precise high resolution gold standard information issued from official channels (which given the emergency nature of the studied phenomena might not exist in most of the cases). In order to validate the utility of the temporal series of z-score measurements to detect floods and potentially other disasters, an exhaustive benchmark against several datasets should be made. Indeed, there are factors that would need special consideration as the difference in the response depends on cultural traits, geographical characteristics and socio-economic status.

While it is clear that there is a need for further development of these methods and techniques, it must also be recognized that the operational implementation of these methodologies also implies institutional capacities, policy frameworks and technological infrastructure that may not be currently in place within local or national disaster management offices.

Potential angles for future research include further validation by combining information from CDRs with data from other sensors, such as traffic video cameras, or by monitoring the time it takes for CDR indicators to stabilize and return to normal levels, as a potential measurement of the rate of recovery. This could be helpful for planning and contributing to measures of resilience [21]. We could compare this indicator across different areas and understand where protracted support may be required. In addition, it would be interesting to combine this passive analysis with actively solicited input from disaster-affected communities when feasible, *e.g.,* by conducing live or automated phone surveys that yield information on outcomes -- health, food security, etc. [22]. In sum, the work presented in this paper is small example of a how a public-private partnership could add value to humanitarian response, working always with anonymized data at an aggregated level to eliminate risks to privacy and be in full compliance with national data protection regulations.

## ACKNOWLEDGEMENTS

We would like to thank the Global Pulse team in New York and the Vulnerability Analysis and Mapping team from WFP in Rome. We thank Ania Calderon and Guillermo Ruiz de Teresa from the Presidencia de la República de México for their support of the project. Thanks to the following research groups from the Universidad Politecnica de Madrid for their contribution: Biomedical Image Technologies lab, Complex Systems group, Earthquake Engineering group and Innovation Centre of Technologies for Human Development.

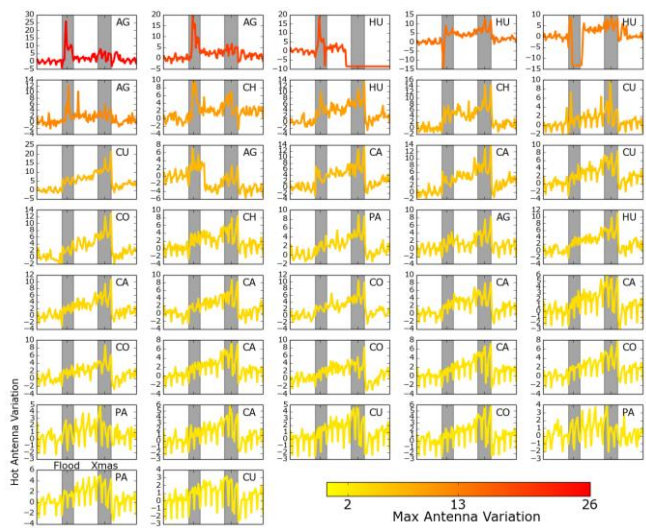

**Appendix-Figure**: Time series of BTS variation per BTS. The gray stripes indicate the floods and Christmas where the metric spikes. The top-right labels in each chart indicate the municipality where the BTS is located (see Fig.6). Charts have been ordered and colored according to the maximum value of the metric for each BTS within the floods segment (hot map from the smallest to the highest variation).

**Appendix-Video**: Time-lapse of the Tabasco impact map (https://www.youtube.com/watch?v=0str5UXDQEU)
The video displays the absolute value of the BTS variation metric from Oct'09 to Jan'10 as in the temporal series. Each antenna is represented by a circle with color and size proportional to the daily metric value. The segmented flooded area has been colored in light blue.